\documentclass[12pt,a4paper,twoside]{article}
\newenvironment{changemargin}[2]{%
\begin{list}{}{%
\setlength{\leftmargin}{#1}%
\setlength{\rightmargin}{#2}%
}%
\item[]}
{\end{list}}
\usepackage{graphicx}
\usepackage{tablefootnote}
\usepackage{color}
\usepackage{lscape}
\usepackage{hyperref}
\usepackage{epstopdf}
\usepackage{lscape}
\usepackage{amsmath}
\usepackage{hyperref}
\usepackage{amsmath}
\usepackage{setspace}
\begin{document}
\baselineskip=0.30in
{\bf \LARGE
\begin{changemargin}{-1.2cm}{0.5cm}
\begin{center}
{Nonrelativistic molecular models under external magnetic and AB flux fields}
\end{center}
\end{changemargin}}

\vspace{4mm}
\begin{center}
{\Large{\bf Sameer M. Ikhdair $^a$$^{,}$}}\footnote{\scriptsize E-mail:~  sameer.ikhdair@najah.edu;~ sikhdair@gmail.com.}\Large{\bf ,}{\Large{\bf Babatunde J. Falaye $^b$$^{,}$$^\dag$$^{,}$}}\footnote{\scriptsize E-mail:~ fbjames11@physicist.net;~ babatunde.falaye@fulafia.edu.ng\\ $\dag$ Corresponding Author Tel. no: +2348103950870 }  \Large{\bf and} {\Large{\bf Majid Hamzavi $^c$$^{,}$}}\footnote{\scriptsize E-mail:~ majid.hamzavi@gmail.com}
\end{center}
{\small
\begin{center}
{\it $^\textbf{a}$Department of Physics, Faculty of Science, an-Najah National University, New campus, P. O. Box 7, Nablus, West Bank, Palestine.}
{\it $^\textbf{b}$Applied Theoretical Physics Division, Department of Physics, Federal University Lafia,  P. M. B. 146, Lafia, Nassarawa State Nigeria.}
{\it $^\textbf{c}$Department of Physics, University of Zanjan, Zanjan, Iran.}
\end{center}}

\begin{center}
Annals of Physics (2014) 10.1016/j.aop.2014.11.017\\
available at: http://www.sciencedirect.com/science/article/pii/S0003491614003376?np=y
\end{center}

\begin{abstract}
\noindent
By using the wave function ansatz method, we study the energy eigenvalues and wave function for any arbitrary $m$-state in two-dimensional Schr\"{o}dinger wave equation with various power interaction potentials in constant magnetic and Aharonov-Bohm (AB) flux fields perpendicular to the plane where the interacting particles are confined. We calculate the energy levels of some diatomic molecules in the presence and absence of external magnetic and AB flux fields using different potential models.  We found that the effect of the Aharonov-Bohm field is much as it creates a wider shift for $m\neq 0$ and its influence on $m=0$ states is found to be greater than that of the magnetic field. To show the accuracy of the
present model, a comparison is made with those ones obtained in the absence of external fields. An extension to 3-dimensional quantum system have also been presented.
\end{abstract}


PACs: 03.65.-w; 03.65.Fd; 03.65.Ge; 71.20.Nr; 73.61.Ey

\section{Introduction}

It is commonly known that the anharmonic and harmonic oscillator potentials play an important role in the history of molecular and quantum chemistry and have also been used to describe the molecular structure and molecules \cite{SM1,SM2}. Also, the Kratzer-type potential describing the molecular vibrations is important in studying the dynamical variables of diatomic molecules \cite{SM3}. This potential have a wide applications in various fields of physics and chemistry such as molecular physics, solid-state physics, chemical physics, quantum chemistry, the molecular dynamics study of linear diatomic molecules and the theoretical works on the spectral properties of a diatomic molecule system \cite{SM4}. Therefore, we found that it is necessary to study the exact bound state solutions of the two-dimensional (2D) solution of the Schr\"{o}dinger equation for these potentials under the influence of external magnetic and Aharonov-Bohm fields.

The 2D hydrogen model was treated as an atomic spectroscopy and used as a simplified model of the ionization process of the highly excited 3D hydrogen atom by circular-polarized microwaves \cite{SM5}. The field-free relativistic Coulomb interaction has been studied by many authors by using various techniques \cite{SM6,SM7,SM8}. The nonrelativistic H-like atom under the influence of magnetic field has been the subject of study over the past years \cite{SM9,SM10,SM11}. In the presence of a low magnetic field, the quasi-classical solution of the Dirac equation has been obtained by factorization method \cite{SM12}. In the framework of the variational method, the ground-state Dirac energies and relativistic spinless lowest few states have been calculated for arbitrary strength values of magnetic field \cite{SM13,SM14,SM15}. The Klein-Gordon wave equation was solved exactly for particular values of magnetic field in which the wave function can be expressed in closed analytical form \cite{SM16}. The polynomial solutions of the Schr\"{o}dinger equation was obtained for the ground-state and a few first excited states of 2D hydrogenic atoms for particular values of the magnetic field strength perpendicular to the plane of transversal motion of the electron using a relativistic wave function \cite{SM17}. Recently, within the framework of power-series solutions, the Klein-Gordon and Dirac equations have been solved for the 2D hydrogen-like systems when an arbitrary external magnetic field is applied \cite{SM18}. For particular values of magnetic field $B,$ it is found that the exact polynomial solutions can be found using the well-known methods in literature \cite{SM16,SM19,SM20,SM21,SM22,SM23}.

Recently, the spectral properties in a 2D charged particle (electron or hole) confined by a PHO potential in the presence of external strong uniform magnetic field $\overrightarrow{B}$ along the $z$ direction and Aharonov-Bohm (AB) flux field created by a solenoid have been studied. The Schr\"{o}dinger equation is solved exactly for its bound states (energy spectrum and wave functions) \cite{SM24,SM25}. So, it is natural that the relativistic effects for a charged particle under the action of this potential could become important, especially for a strong coupling. Within this annals, we have also studied the exact analytical bound state energy eigenvalues and normalized wave functions of the spinless relativistic equation with equal scalar and vector pseudo-harmonic interaction under the effect of external uniform magnetic field and AB flux field in the framework of the NU method \cite{SM26}. Robnik and Romanovski \cite{SM27} studied the Schr\"{o}dinger equation of the hydrogen atom in a strong magnetic field in 2D. Setare and Hatami \cite{SM28} considered the exact solutions of the Dirac equation for an electron in a magnetic field with shape invariant method. Villalba \cite{SM29} analyzed the relativistic Dirac electron in the presence of a combination of a Coulomb field, a $1/r$ scalar potential as well as the Dirac magnetic monopole and an Aharonov-Bohm potential using the algebraic method of separation of variables. Schmid-Burgh and Gail \cite{SM30} solved the stationary Dirac equation for an electron embedded in a uniform magnetic field and in an electrostatic potential periodic along the field lines. The eigenvalues and width of the lowest band gap are calculated. Ko\'{s}cik and Okopi\'{n}ska have studied quasi exact solutions for two interacting
electrons via Coulombic force and confined in an anisotropic harmonic oscillator in 2D anisotropic dot \cite{SM31}. An ansatz for wave function has been applied to obtain the D-dimensional solutions of radial Schr\"{o}dinger equation with some anharmonic potentials \cite{SM32}.

Recently, the 2D solution of the Schr\"{o}dinger equation for the Kratzer potential with and without the presence of a constant magnetic field has been studied for the first time within the framework of the asymptotic iteration method (AIM) \cite{SM33}. Effect of constant magnetic field on the energy eigenvalues of a particle moving under the Kratzer potential was precisely presented by Aygun et al. \cite{SM33}. Furthermore, we have studied the spectral properties of the quantum dots by solving the Schr\"{o}dinger equation for the anharmonic oscillator potential and superposition of pseudoharmonic-linear-Coulomb potential in presence/absence of external uniform magnetic and AB flux fields in the framework of the Nikiforov-Uvarov (NU) method and analytical exact iteration method (AEIM) \cite{SM34}. Also, we have studied the effect of these external fields on the energy states and wave functions of a spinless particle with the Cornell (linear-plus-Coulomb) and Killingbeck (Cornell-plus-harmonic oscillator) potentials, respectively \cite{SM35}. Very recently, we solved the Dirac equation for the anharmonic oscillator potential under the influence of the external magnetic and AB flux fields and obtained the relativistic energy states \cite{SM36}. The two-dimensional solution of the spinless Klein–Gordon (KG) equation for scalar-vector harmonic oscillator potentials with and without the presence of constant perpendicular magnetic and Aharonov–Bohm (AB) flux fields has been studied within the asymptotic function analysis and NU method \cite{JAM4} and via Laplace transform method \cite{JAM5}.

The aim of this work is to extend the works in \cite{SM34,SM35,SM36} and solve the Schr\"{o}dinger equation for a general form of potential models in the presence of external magnetic and AB flux fields. This work can be considered as a unified treatment in studying some molecular potentials which find their direct application in molecular physics and chemistry like the diatomic molecules. By using the quasi-exact solutions, we carry out detailed exact 2D and 3D energetic spectrum and wave functions of the Schr\"{o}dinger equation with a general form of potential models including the well-known molecular interactions in the presence of two external magnetic and Aharonov-Bohm (AB) flux fields. For example, our investigation includes several molecular potentials used to study diatomic molecules such as the pseudo-harmonic interaction, harmonic oscillator, the generalized Kratzer potential, modified Kratrzer potential and Mie-type potential.

The outline of our paper is as follows. In section 2, we give the quasi exact solutions of the 2D Schr\"{o}dinger equation with a general scalar radial potential form $V(r)$ and vector potential $\overrightarrow{A}$ under the influence of external magnetic and AB flux fields. In section 3, the calculations within the quasi exact solution are obtained for various molecular potentials to obtain the closed analytical energy spectrum and wave functions under external fields and the comparison with exact results is given when fields become zero. Finally, the paper ends with concluding remarks in section 4.

\section{Exact Solution to a General Potential Form under External Fields}

Consider a 2D charged particles with charge $e$ and effective mass $\mu $
interacting via a radially symmetrical interaction potential $V(r,\phi )$
under the influence of external uniform magnetic field, $\overrightarrow{B}=B%
\widehat{z}$ and an AB flux field, applied simultaneously. Since we are
taking the Hamiltonian in 2D, its adequate to study the system in polar
coordinates $(r,\phi )$ within the plane. Hence, the Schr\"{o}dinger
equation with a vector potential $\overrightarrow{A}$ \ and repulsive
interaction potential $V(r,\phi )$ can be written as 
\begin{equation}
\left[ \frac{1}{2\mu }\left( \overrightarrow{p}+\frac{e}{c}\overrightarrow{A}%
\right) ^{2}+V(r)\right] \psi (r,\phi )=E\psi (r,\phi ),
\end{equation}%
with the 2D cylindrical wave function 
\begin{equation}
\psi (r,\phi )=\frac{1}{\sqrt{2\pi }}e^{im\phi }u(r),\text{ }m=0,\pm 1,\pm
2,\ldots ,
\end{equation}%
and $m$ is the magnetic quantum number (the eigenvalue of angular momentum).
For the solution of Eq. (1), we take the potential $V(r,\phi )$ as radial
scalar power form \cite{SM23,SM24} 
\begin{equation}
V(r)=ar^{2}+br+d-\frac{g}{r}+\frac{k}{r^{2}},
\end{equation}%
and the vector potential $\overrightarrow{A}$ may be represented as a sum of
two terms, $\overrightarrow{A}=\overrightarrow{A}_{1}+\overrightarrow{A}_{2}$
such that $\overrightarrow{\nabla }\times \overrightarrow{A}_{1}=%
\overrightarrow{B}$ and $\overrightarrow{\nabla }\times \overrightarrow{A}%
_{2}=0,$ where $\overrightarrow{B}$ $=B\widehat{z}$ is the applied magnetic
field, and $\overrightarrow{A}_{2}$ describes the additional magnetic flux $%
\Phi _{AB}$ created by a solenoid. Hence, the vector potentials in the
symmetric gauge have the following azimuthal components \cite{SM25} 
\begin{equation}
\overrightarrow{A}_{1}=\frac{1}{2}\overrightarrow{B}\times \overrightarrow{r}%
=\frac{Br}{2}\widehat{\phi },\text{ }\overrightarrow{A}_{2}=\frac{\Phi _{AB}%
}{2\pi r}\widehat{\phi },\text{ }\overrightarrow{A}=\left( \frac{Br}{2}+%
\frac{\Phi _{AB}}{2\pi r}\right) \widehat{\phi }.
\end{equation}%
Substituting the wave function (2) into the Schr\"{o}dinger equation (1), we
obtain a second-order differential equation satisfying the radial part of
the wave function; namely $u(r),$%
\begin{equation}
\frac{d^{2}u(r)}{dr^{2}}+\frac{1}{r}\frac{du(r)}{dr}+\left( -\varepsilon
^{2}-\frac{\beta ^{2}}{r^{2}}-\gamma ^{2}r^{2}+\frac{g^{\prime }}{r}%
-b^{\prime }r\right) u(r)=0,
\end{equation}%
with 
\begin{subequations}
\begin{equation}
-\varepsilon ^{2}=E^{\prime }-d^{\prime }-\frac{\mu \omega _{c}}{\hbar }%
m^{\prime },
\end{equation}%
\begin{equation}
\beta ^{2}=k^{\prime }+m^{\prime 2},
\end{equation}%
\begin{equation}
\gamma ^{2}=a^{\prime }+\left( \frac{\mu \omega _{c}}{2\hbar }\right) ^{2},
\end{equation}%
\begin{equation}
g^{\prime }=\frac{2\mu }{\hbar ^{2}}g,\text{ }b^{\prime }=\frac{2\mu }{\hbar
^{2}}b,\text{ }E^{\prime }=\frac{2\mu }{\hbar ^{2}}E,\text{ }d^{\prime }=%
\frac{2\mu }{\hbar ^{2}}d,\text{ }k^{\prime }=\frac{2\mu }{\hbar ^{2}}k,%
\text{ }a^{\prime }=\frac{2\mu }{\hbar ^{2}}a.
\end{equation}%
and 
\end{subequations}
\begin{equation}
\xi =\frac{\Phi _{AB}}{\Phi _{0}},\text{ }\Phi _{0}=\frac{hc}{e},\text{ }%
\omega _{c}=\frac{eB}{\mu c},\text{ }m^{\prime }=m+\xi .
\end{equation}%
Note that $\xi $ is taken as integer with the flux quantum $\Phi _{0}$ and $%
\omega _{c}$ is the cyclotron frequency. The magnetic quantum number (the
eigenvalue of angular momentum) $m$ relates to the new quantum number $%
\left\vert \beta \right\vert $ [Eq. (6b)]. For this system, only two
independent integer quantum numbers are required. In addition, the radial
wave function $u(r)$ must satisfy the asymptotic behaviors,\textit{\ }that
is,$\ $\ $u(0)\rightarrow 0$ and $u(\infty )\rightarrow 0.$ To make the
coefficient of the first derivative vanish in Eq. (5), we may define radial
wave function $R(r)$ by means of the equation 
\begin{equation}
u(r)=r^{-1/2}R(r),
\end{equation}%
which will lead to the radial wave function $R(r)$ satisfying%
\begin{equation}
\frac{d^{2}R(r)}{dr^{2}}+\left( -\varepsilon ^{2}-\gamma ^{2}r^{2}+\frac{%
g^{\prime }}{r}-b^{\prime }r-\frac{\left( \beta ^{2}-1/4\right) }{r^{2}}%
\right) R(r)=0.
\end{equation}%
To solve the above radial differential equation for $\beta \neq 0,$ we first
investigate the asymptotic behavior of $R(r).$ First inspection of Eq. (9)
shows that when $r$ approaches $0,$ the asymptotic solution $R_{0}(r)$ of
Eq. (9) satisfies the differential equation%
\begin{equation}
\frac{d^{2}R_{0}(r)}{dr^{2}}-\frac{\left( \beta ^{2}-1/4\right) }{r^{2}}%
R_{0}(r)=0,
\end{equation}%
which assumes the solution $R_{0}(r)=c_{1}r^{\beta +1/2}+c_{2}r^{-\left(
\beta +1/2\right) },$ where $c_{1}$ and $c_{2}$ are two constants. The term $%
r^{-\left( \beta +1/2\right) }$ is not a satisfactory solution because it
becomes infinite as $r\rightarrow 0$ but the term $r^{\beta +1/2}$ is well
behaved. Meanwhile, as $r$ approaches $\infty ,$ the asymptotic solution $%
R_{\infty }(r)$ of Eq. (9) gives the differential equation%
\begin{equation}
\frac{d^{2}R_{\infty }(r)}{dr^{2}}-\left( b^{\prime }r+\gamma
^{2}r^{2}\right) R_{\infty }(r)=0,
\end{equation}%
which yields the solution $R_{\infty }(r)=c_{3}\exp \left[ g(p,q,r)\right] ,$
where $g(b^{\prime },\gamma ^{2},r)=-b^{\prime }r-\gamma ^{2}r^{2}$ is an
acceptable physical solution since the solution becomes finite as $%
r\rightarrow \infty .$ Consequently, this asymptotic behavior of $R(r)$
suggests we take an ansatz for the radial wave function \cite{SM32}%
\begin{equation}
R(r)=r^{\beta +1/2}\exp \left[ g(p,q,r)\right] F(r),
\end{equation}%
with 
\begin{subequations}
\begin{equation}
g(p,q,r)=qr+\frac{1}{2}pr^{2},\text{ }q<0,\text{ }p<0
\end{equation}%
\begin{equation}
F(r)=\sum_{n=0}a_{n}r^{n}.
\end{equation}%
Substituting Eq. (12) into Eq. (9) and equatimg the coefficients of $%
r^{n+\beta +1/2}$ to zero, we can obtain the following recurrence relation, 
\end{subequations}
\begin{equation}
A_{n}a_{n}+B_{n+1}a_{n+1}+C_{n+2}a_{n+2}=0,
\end{equation}%
where 
\begin{subequations}
\begin{equation}
A_{n}=-\varepsilon ^{2}+q^{2}+2p\left( n+\beta +1\right) ,
\end{equation}%
\begin{equation}
B_{n}=g^{\prime }+q\left( 2n+2\beta +1\right) ,
\end{equation}%
\begin{equation}
C_{n}=n\left( n+2\beta \right) ,
\end{equation}%
and 
\end{subequations}
\begin{subequations}
\begin{equation}
p^{2}=\gamma ^{2},
\end{equation}%
\begin{equation}
2pq=b^{\prime }.
\end{equation}%
It is shown from Eq. (16) that the values of the parameters for $g(p,q,r)$
can be evaluated as 
\end{subequations}
\begin{equation}
p=\pm \gamma ,\text{ }q=\frac{b^{\prime }}{2p}.
\end{equation}%
To retain a well-behaved solution at the origin and at $\infty ,$ we choose $%
p=-\gamma $ which enables us to write $q=-b^{\prime }/(2\gamma ).$ On the
other hand, for a given $s,$ if $a_{s}\neq 0$ but $a_{s+1}=a_{s+2}=a_{s+3}=%
\cdots =0,$ where $s$ is the degree of polynomial $(s=0,1,2,\cdots ),$ we
then impose the condition $A_{s}=0$ from Eq. (14), i.e., from which we can
obtain the eigenvalue equation 
\begin{equation}
\varepsilon ^{2}=q^{2}+2p\left( s+\beta +1\right) ,
\end{equation}%
which establishes the relationship between energies, parameters of the
potential and the two fields strength $B$ and $\xi $. Substituting the
values of the parameters defined in (6) into (18) and by using $p$ and $q$
obtained from (17) we may write the energy eigenvalues of the power
potential as%
\begin{equation*}
E_{nm^{\prime }}=d+\frac{1}{2}\hbar m^{\prime }\omega _{c}+\frac{\hbar ^{2}}{%
2\mu }\sqrt{\frac{8\mu a}{\hbar ^{2}}+\left( \frac{\mu \omega _{c}}{\hbar }%
\right) ^{2}}\left( s+1+\sqrt{\frac{2\mu k}{\hbar ^{2}}+m^{\prime 2}}\right)
\end{equation*}%
\begin{equation*}
-\frac{2\mu b^{2}/\hbar ^{2}}{\left[ \frac{8\mu a}{\hbar ^{2}}+\left( \frac{%
\mu \omega _{c}}{\hbar }\right) ^{2}\right] }.
\end{equation*}%
We follow Refs. \cite{SM32,SM33}, the parameters $A_{n},$ $B_{n}$ and $C_{n}$ must
satisfy the determinant relation for a nontrivial solution:%
\begin{equation}
\det \left\vert 
\begin{array}{cccccc}
B_{0} & C_{1} & \cdots & \cdots & \cdots & 0 \\ 
A_{0} & B_{1} & C_{2} & \cdots & \cdots & 0 \\ 
\vdots & \vdots & \vdots & \ddots & \vdots & \vdots \\ 
0 & 0 & 0 & 0 & A_{s-1} & B_{s}%
\end{array}%
\right\vert =0.
\end{equation}%
Note that Eq. (19) leads to the restriction for the flux quantum $\Phi _{0}$
and the cyclotron frequency $\omega _{c}$ if all other parameters of the
model are fixed. However, Eq. (18) allows for calculation the energies for
these particular $\Phi _{0}$ and $\omega _{c}.$

As an example, the exact solution for $s=0,1$ are demonstrated below.

(1) For the simplest polynomial solution $(s=0),$ we can obtain from Eq.
(18) 
\begin{equation}
\varepsilon _{0}^{2}=-2\gamma \left( \beta +1\right) +\frac{\mu ^{2}b^{2}}{%
\hbar ^{4}\gamma ^{2}}.
\end{equation}%
On the other hand, it is shown from Eq. (19) that $B_{0}=0,$ which leads to
the following restriction on the parameters of the potential and $\beta $: 
\begin{equation}
q=-\frac{g^{\prime }}{\left( 2\beta +1\right) },
\end{equation}%
and the enrgy formula can be obtained by substituting Eqs. (6b), (6d) and
(17) into Eq. (21) 
\begin{equation}
b=\frac{\gamma }{\left( \frac{1}{2}+\sqrt{\frac{2\mu }{\hbar ^{2}}%
k+m^{\prime 2}}\right) }g.
\end{equation}%
The explicit form of the ground state energy is 
\begin{equation*}
E_{0}^{(m^{\prime })}=d+\frac{1}{2}\hbar m^{\prime }\omega _{c}+\frac{\hbar
^{2}}{\mu }\sqrt{\left( \frac{\mu \omega _{c}}{2\hbar }\right) ^{2}+\frac{%
2\mu }{\hbar ^{2}}a}\left( 1+\sqrt{m^{\prime 2}+\frac{2\mu }{\hbar ^{2}}k}%
\right)
\end{equation*}

\begin{equation}
-\frac{\mu }{2\hbar ^{2}}\frac{b^{2}}{\left( \frac{2\mu }{\hbar ^{2}}%
a+\left( \frac{\mu \omega _{c}}{2\hbar }\right) ^{2}\right) }.
\end{equation}%
The eigenfunction for $s=0$ can be written as 
\begin{equation}
\psi ^{0}(r,\phi )=\frac{1}{\sqrt{2\pi }}a_{0}r^{\sqrt{m^{\prime 2}+\frac{%
2\mu }{\hbar ^{2}}k}}\exp \left[ -\frac{1}{2}\left( \frac{2\mu b}{\hbar
^{2}\left( \frac{2\mu }{\hbar ^{2}}a+\left( \frac{\mu \omega _{c}}{2\hbar }%
\right) ^{2}\right) ^{1/2}}r+\gamma r^{2}\right) \right] e^{im\phi },
\end{equation}%
where $a_{0}$ is a expansion constant.

(2) When $s=1,$ the eigenvalue from Eq. (18) becomes 
\begin{equation*}
\varepsilon _{1}^{2}=-2\gamma \left( \beta +2\right) +\frac{\mu ^{2}b^{2}}{%
\hbar ^{4}\gamma ^{2}},
\end{equation*}%
and with the aid of Eq. (6) it reads%
\begin{equation*}
E_{1}^{(m^{\prime })}=d+\frac{1}{2}\hbar m^{\prime }\omega _{c}+\frac{\hbar
^{2}}{\mu }\sqrt{\left( \frac{\mu \omega _{c}}{2\hbar }\right) ^{2}+\frac{%
2\mu }{\hbar ^{2}}a}\left( 2+\sqrt{m^{\prime 2}+\frac{2\mu }{\hbar ^{2}}k}%
\right)
\end{equation*}

\begin{equation}
-\frac{\mu }{2\hbar ^{2}}\frac{b^{2}}{\left( \frac{2\mu }{\hbar ^{2}}%
a+\left( \frac{\mu \omega _{c}}{2\hbar }\right) ^{2}\right) }.
\end{equation}%
Moreover,it is shown from Eq. (19) that $B_{0}B_{1}-A_{0}C_{1}=0$ which
provides the restriction on the parameters: 
\begin{equation*}
\left[ g^{\prime }-\frac{b^{\prime }}{2\gamma }\left( 2\beta +1\right) %
\right] \left[ g^{\prime }-\frac{b^{\prime }}{2\gamma }\left( 2\beta
+3\right) \right] =2\gamma \left( 2\beta +1\right) ,
\end{equation*}%
or alternatively%
\begin{equation}
g=\frac{b}{\gamma }\left( \beta +1\right) \pm \frac{\hbar ^{2}}{4\mu }\sqrt{%
\frac{\mu ^{2}}{\hbar ^{4}}\frac{b^{2}}{\gamma ^{2}}+2\gamma \left( 2\beta
+1\right) }.
\end{equation}%
The corresponding wave function can be written as%
\begin{equation}
\psi ^{1}(r,\phi )=\frac{1}{\sqrt{2\pi }}\left( a_{0}+a_{1}r\right) r^{\sqrt{%
m^{\prime 2}+\frac{2\mu }{\hbar ^{2}}k}}\exp \left[ -\frac{1}{2}\left( \frac{%
2\mu b}{\hbar ^{2}\left( \frac{2\mu }{\hbar ^{2}}a+\left( \frac{\mu \omega
_{c}}{2\hbar }\right) ^{2}\right) ^{1/2}}r+\gamma r^{2}\right) \right]
e^{im\phi },
\end{equation}%
where $a_{0}$ and $a_{1}$ are expansion constants.

Following this way, we can obtain a class of exact solutions through taking $%
s=0,1,2,\cdots $ etc. We obtain the energy levels from Eq. (18) with the aid
of Eq. (6) as%
\begin{equation*}
E_{nm^{\prime }}=d+\frac{1}{2}\hbar m^{\prime }\omega _{c}+\frac{\hbar ^{2}}{%
\mu }\sqrt{\left( \frac{\mu \omega _{c}}{2\hbar }\right) ^{2}+\frac{2\mu }{%
\hbar ^{2}}a}\left( n+1+\sqrt{m^{\prime 2}+\frac{2\mu }{\hbar ^{2}}k}\right)
\end{equation*}

\begin{equation}
-\frac{\mu }{2\hbar ^{2}}\frac{b^{2}}{\left( \frac{2\mu }{\hbar ^{2}}%
a+\left( \frac{\mu \omega _{c}}{2\hbar }\right) ^{2}\right) },
\end{equation}%
and the wave function can be obtained from Eqs. (2), (8), (12) and (13) as%
\begin{equation*}
\psi (r,\phi )=\frac{1}{\sqrt{2\pi }}e^{im\phi }\exp \left[ -\frac{1}{2}%
\left( \frac{2\mu b}{\hbar ^{2}\left( \frac{2\mu }{\hbar ^{2}}a+\left( \frac{%
\mu \omega _{c}}{2\hbar }\right) ^{2}\right) ^{1/2}}r+\gamma r^{2}\right) %
\right]
\end{equation*}%
\begin{equation}
\times \sum_{n=0}a_{n}r^{n+\sqrt{m^{\prime 2}+\frac{2\mu }{\hbar ^{2}}k}},
\end{equation}%
where $a_{n}$ $(n=0,1,2,\cdots s)$ are expansion constants. We have one
set of quantum numbers $(n,m,\beta )$ for particle. Therefore, expression
(24) for the energy levels of the electron may be readily used for studying
the thermodynamics properties of quantum structures when the magnetic field
turns on or off.

There is a corresponding relationship between 2D and 3D obtained by making a
replacement $m=l+1/2$. Therefore, the bound state energy levels and wave
function in 3D have the new forms:%
\begin{equation*}
E_{nl}=d+\frac{\hbar ^{2}}{\mu }\sqrt{\frac{2\mu }{\hbar ^{2}}a}\left( n+1+%
\sqrt{\left( l+\frac{1}{2}\right) ^{2}+\frac{2\mu }{\hbar ^{2}}k}\right)
\end{equation*}

\begin{equation}
-\frac{\mu }{2\hbar ^{2}}\frac{g^{2}}{\left( n+\frac{1}{2}+\sqrt{\left( l+%
\frac{1}{2}\right) ^{2}+\frac{2\mu }{\hbar ^{2}}k}\right) ^{2}},
\end{equation}%
\begin{equation*}
\psi (r,\theta ,\phi )=\exp \left[ -\frac{1}{2}\left( \frac{2\mu g}{\left( n+%
\frac{1}{2}+\sqrt{\left( l+\frac{1}{2}\right) ^{2}+\frac{2\mu }{\hbar ^{2}}k}%
\right) \hbar ^{2}}r+\gamma r^{2}\right) \right]
\end{equation*}%
\begin{equation}
\times \sum_{n=0}a_{n}r^{n+\sqrt{\left( l+\frac{1}{2}\right) ^{2}+\frac{2\mu 
}{\hbar ^{2}}k}}Y_{lm}(\theta ,\phi ),
\end{equation}%
respectively, where $Y_{lm}(\theta ,\phi )$ is the spherical harmonic
function with angular momentum quantum numbers being $l$ and $m.$

\section{Applications}

The structure of molecules is more complex than atoms, however, we may
assume that the ratio of the molecular mass to electron mass to be large,
which implies that the energy associated with the motion of the nuclei is
much smaller than that associated with the motion of the electrons around
the nuclei. Hence, this simplification is the basis of all molecular
approximations. It is a good approximation to assume the nuclei to have
stable equilibrium arrangement between collapsed structure and dispersed
structure while studying the electronic motion. This nuclear structure can
be devided into translations and rotations of the quasi-rigid equilibrium
arrangement and internal vibrations of the nuclei about equilibrium, the
translational motion does not give rise to non-classical features. Thus, the
molecular energy levels and wave functions are classified into electronic,
vibrational and rotational types \cite{SM37}.

The nuclei in diatomic molecules in general have masses $m_{1}$ and $m_{2}$
and their relative position vector $\overrightarrow{r}$ and cylindrical
coordinates $r,\phi ,z$ or spherical coordinates $r,\theta ,\phi .$

We apply our results obtained in the previous section to a class of
molecular potentials which includes the pseudo-harmonic interaction, the
harmonic oscillator, the generalized Kratzer potential, the modified Kratzer
potential and the Mie-type potential. Such potentials are reasonably behaved
for both small and large internuclear separations. This set of potential
models provides a reasonable description of the rotating diatomic molecules
and also will be useful in discussing long amplitude vibrations in large
molecules. We will find the energy levels and wave functions of a rotating
diatomic molecules in 2D space when applying the magnetic field and also 3D
space without magnetic field.

\subsection{The pseudoharmonic oscillator}

The pseudoharmonic oscillator is an important model not only in classical physics, but also in quantum physics. In non-relativistic quantum mechanics, many authors have adequately studied the problem of the pseudoharmonic oscillator \cite{SM36,SM37,SM38,SM39}. It has been studied in 3D \cite{SM36} using the polynomial solution method, in 2D \cite{SM39} using the Nikiforov-Uvarov method and in $D$-dimensions \cite{SM21} using an ansatz for the wave function. It has the following form \cite{SM38,SM39,SM40,SM41} 
\begin{equation}
V(r)=D_{e}\left( \frac{r}{r_{e}}-\frac{r_{e}}{r}\right) ^{2},
\end{equation}%
which can be simply rewritten in the form of isotropic harmonic oscillator
plus square inverse potential 
\begin{equation}
V(r)=ar^{2}+\frac{k}{r^{2}}+d,\text{ }a,k>0,
\end{equation}%
where $a=D_{e}r_{e}^{-2},$ $k=D_{e}r_{e}^{2}$ and $d=-2D_{e}.$ Therefore,
the energy formula and wave function from (28) and (29) become%
\begin{equation}
E_{nm^{\prime }}=-2D_{e}+\hbar \Omega \left( n+\frac{\left\vert \beta
\right\vert +1}{2}\right) +\frac{1}{2}\hbar m^{\prime }\omega _{c},\text{ }%
\Omega =\sqrt{\omega _{c}^{2}+4\omega _{D}^{2}},
\end{equation}%
\begin{equation}
\psi (r,\phi )=\frac{1}{\sqrt{2\pi }}e^{im\phi }\exp \left[ -\frac{1}{2}%
\sqrt{\left( \frac{\mu \omega _{c}}{2\hbar }\right) ^{2}+\frac{2\mu D_{e}}{%
\hbar ^{2}r_{e}^{2}}}\text{ }r^{2}\right] \sum_{n=0}a_{n}r^{2n+\sqrt{%
m^{\prime 2}+\frac{2\mu D_{e}r_{e}^{2}}{\hbar ^{2}}}\text{ }},
\end{equation}%
where $\left\vert \beta \right\vert =\sqrt{m^{\prime 2}+\frac{2\mu
D_{e}r_{e}^{2}}{\hbar ^{2}}}$ and $\omega _{D}=\sqrt{2D_{e}/\mu r_{e}^{2}}.$
This is identical to Refs. \cite{SM24,SM26,SM41} for the nonrelativistic
pseudo-harmonic in an external uniform magnetic and AB flux fields. Further,
in the absence of the external fields, i.e., $B=\Phi _{AB}=0,$ we further
obtain,

\begin{equation}
E_{nm}=-2D_{e}+\frac{\hbar }{r_{e}}\sqrt{\frac{2D_{e}}{\mu }}\left( 2n+1+%
\sqrt{m^{2}+\frac{2\mu D_{e}r_{e}^{2}}{\hbar ^{2}}}\right) ,
\end{equation}%
\begin{equation}
\psi (r,\phi )=B_{nm}\frac{1}{\sqrt{2\pi }}e^{im\phi }\exp \left[ -\frac{1}{2%
}\sqrt{\frac{2\mu D_{e}}{\hbar ^{2}r_{e}^{2}}}\text{ }r^{2}\right]
\sum_{n=0}a_{n}r^{2n+\sqrt{m^{2}+\frac{2\mu D_{e}r_{e}^{2}}{\hbar ^{2}}}%
\text{ }},
\end{equation}%
and reduces to 3D space if one replaces $m$ by $l+1/2$ as%
\begin{equation}
E_{nl}=-2D_{e}+\frac{\hbar }{r_{e}}\sqrt{\frac{2D_{e}}{\mu }}\left( 2n+1+%
\sqrt{\left( l+\frac{1}{2}\right) ^{2}+\frac{2\mu D_{e}r_{e}^{2}}{\hbar ^{2}}%
}\right) ,
\end{equation}%
which is identical to Eq. (15) of Ref. \cite{SM38}.

In our numerical work, we estimate the non-relativistic binding energy levels of the pseudo-harmonic potential for NO, CO, N$_{2}$ and CH diatomic molecules with various vibrational $n$ and rotational $l$ quantum numbers, using Eq. (38) in 3D space, with the potential parameter values given in Ref. \cite{SM37}. We display our results in Table 1 and compare them with Ref. \cite{SM38}. Further, the energy levels of these diatomic molecules for any arbitrary quantum numbers $n$ and $m$ obtained through Eq. (34) under the external magnetic field $B$ and AB flux field $\xi $ of different strength values. The results are displayed in Table 2 for N$_{2}$ and CH molecules. It is noticed from Table 2 that the magnetic field strength $B\neq 0$ has a very small effect on the energy levels when the magnetic quantum number $m=0.$ However, it has a greater effect on the energy levels when $m\neq 0;$ leading to a significant splitting to the two degenerate states $m=1$ and $m=-1.$ Therefore, we conclude that only the strong magnetic fields have impact on the $m=0.$ However, the magnetic field of any strength has greater effect on the magnetic quantum number $m\neq 0$. When we set $B=0$ and $\xi \neq 0,$ the Aharanov-Bohm field of any strength has greater influence on the the energy levels for any $m$ value. It removes the degeneracy on the $m=1$ and $m=-1$ and splitting them up and down, respectively. The application of a magnetic field of constant strength and at the same time increasing the strength of AB field leads to an interchange in the energy levels for $m=0$ to become the energy level of $m=-1$ and the energy of the state $m=1$ becomes the energy of the state $m=0.$

\subsection{The harmonic oscillator}
For an isotropic harmonic oscillator of angular frequency $\omega ,$ the potential function is given by \cite{SM42}
\begin{equation}
V(r)=\frac{1}{2}\kappa r^{2},\text{ }\kappa =\mu \omega ^{2}.
\end{equation}%
In this case, the energy spectrum formula and wave function become%
\begin{equation}
E_{nm^{\prime }}=\hbar \sqrt{\omega _{c}^{2}+4\omega ^{2}}\left( n+\frac{%
1+m^{\prime }}{2}\right) +\frac{1}{2}\hbar m^{\prime }\omega _{c},
\end{equation}%
\begin{equation}
\psi (r,\phi )=\frac{1}{\sqrt{2\pi }}e^{im\phi }\exp \left[ -\frac{1}{2}%
\sqrt{\left( \frac{\mu \omega _{c}}{2\hbar }\right) ^{2}+\frac{2\mu D_{e}}{%
\hbar ^{2}r_{e}^{2}}}\text{ }r^{2}\right] \sum_{n=0}a_{n}r^{2n+m^{\prime }%
\text{ }},
\end{equation}%
This is identical to Refs. \cite{SM24,SM26} for the nonrelativistic harmonic
oscillator in an external uniform magnetic and AB flux fields. Furthermore,
when the fields $B=\Phi _{AB}=0,$ we can obtain,%
\begin{equation}
E_{nm}=\hbar \omega \left( 2n+1+m\right) ,
\end{equation}%
\begin{equation}
\psi (r,\phi )=\frac{1}{\sqrt{2\pi }}e^{im\phi }\exp \left[ -\frac{\mu
\omega }{2\hbar }\text{ }r^{2}\right] \sum_{n=0}a_{n}r^{2n+m\text{ }}.
\end{equation}%
In 3D, the energy formula becomes $E_{nl}=\hbar \omega \left( 2n+l+\frac{3}{2%
}\right)$

\subsection{Generalized Kratzer Potential}

The generalized Kratzer potential is given by 
\begin{equation}
V(r)=D_{e}\left( \frac{r-r_{e}}{r}\right) ^{2}+\eta ,
\end{equation}
where $r_{e}$ is the the equilibrium separation and $D_{e}$ is the
dissociation energy between diatomic molecules. In the case if $\eta =0,$
this potential reduces to the modified Kratzer potential proposed by Simons
et al \cite{SM43} and by Molski and Konarski \cite{SM44}. The modified Kratzer potential
is usually used in applications on the molecular spectroscopies. Pl\'{\i}va
\cite{SM45} has demonstrated that the dificiencies of rovibrational energy spectrum
based on modified Kratzer potential can be substituted by including
correction terms dependent on the vibrational and rotational quantunm
numbers $n$ and $l,$ respectively. In the case if $\eta =-D_{e},$ this
potential turns to become the Kratzer potential consisting from the
attractive Coulomb potential plus repulsive inverse square potential \cite{SM46}.
Now we want to obtain a 2D energy formula and wave function in the presence
of the generalized Kratzer potential and magnetic field and AB flux field.
So we set the parameters values $a=b=0,$ $d=D_{e}+\eta ,$ $g=2r_{e}D_{e}$
and $k=r_{e}^{2}D_{e}$ and use Eqs. (28) and (29) to obtain%
\begin{equation*}
E_{nm^{\prime }}=\frac{\hbar \omega _{c}}{2}\left[ m^{\prime }+\left( n+1+%
\sqrt{\frac{2\mu r_{e}^{2}D_{e}}{\hbar ^{2}}+m^{\prime }{}^{2}}\right) %
\right]
\end{equation*}%
\begin{equation}
-\frac{2\mu r_{e}^{2}D_{e}^{2}}{\hbar ^{2}}\left( n+\frac{1}{2}+\sqrt{\frac{%
2\mu r_{e}^{2}D_{e}}{\hbar ^{2}}+m^{\prime }{}^{2}}\right) ^{-2}+D_{e}+\eta ,
\end{equation}%
and%
\begin{equation*}
\psi (r,\phi )=\exp \left[ -\frac{2\mu r_{e}D_{e}}{\hbar ^{2}}\left( n+\frac{%
1}{2}+\sqrt{\frac{2\mu r_{e}^{2}D_{e}}{\hbar ^{2}}+m^{\prime }{}^{2}}\right)
^{-1}r-\frac{\mu \omega _{c}}{4\hbar }r^{2}\right]
\end{equation*}%
\begin{equation}
\times \sum_{n=0}a_{n}r^{n+\sqrt{\frac{2\mu r_{e}^{2}D_{e}}{\hbar ^{2}}%
+m^{\prime }{}^{2}}}\frac{1}{\sqrt{2\pi }}e^{im\phi },
\end{equation}%
respectively. Further, when we set $B=0$ and $\Phi _{AB}=0,$ we obtain%
\begin{equation}
E_{nm}=-\frac{2\mu r_{e}^{2}D_{e}^{2}}{\hbar ^{2}}\left( n+\frac{1}{2}+\sqrt{%
\frac{2\mu r_{e}^{2}D_{e}}{\hbar ^{2}}+m^{2}}\right) ^{-2}+D_{e}+\eta ,
\end{equation}%
and%
\begin{equation}
\psi (r,\phi )=\exp \left[ -\frac{2\mu r_{e}D_{e}}{\left( n+\frac{1}{2}+%
\sqrt{\frac{2\mu r_{e}^{2}D_{e}}{\hbar ^{2}}+m^{2}}\right) \hbar ^{2}}r%
\right] \sum_{n=0}a_{n}r^{n+\sqrt{\frac{2\mu r_{e}^{2}D_{e}}{\hbar ^{2}}%
+m^{2}}}\frac{1}{\sqrt{2\pi }}e^{im\phi }.
\end{equation}%
The 3D non-relativistic energy solutions are obtained by setting $m=l+1/2$
where $l$ is the rotational quantum number, in Eq. (47) to obtain%
\begin{equation}
E_{nl}=-\frac{8\mu r_{e}^{2}D_{e}^{2}}{\hbar ^{2}}\left( 1+2n+\sqrt{\frac{%
8\mu r_{e}^{2}D_{e}}{\hbar ^{2}}+\left( 2l+1\right) ^{2}}\right)
^{-2}+D_{e}+\eta ,\text{ }n=0,1,2,\cdots
\end{equation}%
where $l=0,1,2,\cdots $ is the rotational quantum number. Note that Eq. (49)
is identical to Eq. (36) of Ref. \cite{SM47} when $N=3$ and reverts to the
rovibrational energy for the Kratzer potential $($for $\eta =-D_{e})$
studied in \cite{SM48,SM49,SM50} and to energy for the modified Kratzer potential $($for $%
\eta =0)$ studied in \cite{SM51,SM52,SM53}.

In our numerical work, we estimate the non-relativistic binding energy levels of the Kratzer potential for N$_{2}$ and CH diatomic molecules with various vibrational $n$ and rotational $l$ quantum numbers in 3D space, with the potential parameter values given in Ref. \cite{SM37} to see the accuracy of our solution. Our results are displayed in Table 3 and compared with the ones in Ref. \cite{SM37}. Furthermore the energy levels of these diatomic molecules in the presence of external magnetic field $B,$ and AB flux field, $\xi $ of different strength values with various values of quantum numbers $n$ and $m$ obtained through Eq. (45) are presented in Table 4. It is noticed from Table 4 that the magnetic field strength $B\neq 0$ has very small effect on the energy levels when the magnetic quantum number $m=0.$ However, it has greater influence on the energy levels when $m\neq 0;$ it makes splitting to the two degenerate states $m=1$ and $m=-1.$ Therefore, we conclude that only strong magnetic fields have impact on the $m=0.$ But the magnetic of any strength has greater effect on the magnetic quantum number. When $B=0$ and $\xi \neq 0,$ the AB field of any strength has greater influence on the the energy levels for any $m$ value. It removes the degeneracy on the $m=1$ and $m=-1$ and splitting them up and down, respectively. The application of a magnetic field of constant strength and at the same time increasing the strength of AB field leads to an interchange in the energy levels for $m=0$ to become the energy level of $m=-1$ and the energy of the state $m=1$ becomes the energy of the state $m=0.0.$

\subsection{The Mie-type potentials}
This potential has been studied in the $D$ dimensions using the polynomial solution and the ansatz wave function method \cite{SM54,SM55}. The Mie-type potential has the form \cite{SM54}
\begin{equation}
V(r)=D_{e}\left[ \frac{k}{j-k}\left( \frac{r_{e}}{r}\right) ^{j}-\frac{j}{j-k%
}\left( \frac{r_{e}}{r}\right) ^{k}\right] ,\text{ }j>k,
\end{equation}%
where the parameter $D_{e}$ determines the interaction energy between two
atoms in a solid at $r=r_{e}.$ Taking $\ j=2k$ and further setting $k=1,$
the potential reduces to the Coulombic-type form \cite{SM56,SM57,SM58,SM59,SM60}

\begin{equation}
V(r)=2D_{e}\left( \frac{1}{2}\left( \frac{r_{e}}{r}\right) ^{2}-\frac{r_{e}}{%
r}\right) .
\end{equation}%
Now taking $a=b=d=0,$ $g=2r_{e}D_{e}$ and $k=r_{e}^{2}D_{e}$ and using Eqs.
(28) and (29) to obtain%
\begin{equation*}
E_{nm^{\prime }}=\frac{\hbar \omega _{c}}{2}\left[ m^{\prime }+\left( n+1+%
\sqrt{m^{\prime }{}^{2}+\frac{2\mu r_{e}^{2}D_{e}}{\hbar ^{2}}}\right) %
\right]
\end{equation*}%
\begin{equation}
-\frac{2\mu r_{e}^{2}D_{e}^{2}}{\hbar ^{2}}\left( n+\frac{1}{2}+\sqrt{%
m^{\prime }{}^{2}+\frac{2\mu r_{e}^{2}D_{e}}{\hbar ^{2}}}\right) ^{-2},
\end{equation}%
and%
\begin{equation*}
\psi (r,\phi )=\exp \left[ -\frac{2\mu r_{e}D_{e}}{\hbar ^{2}}\left( n+\frac{%
1}{2}+\sqrt{m^{\prime }{}^{2}+\frac{2\mu r_{e}^{2}D_{e}}{\hbar ^{2}}}\right)
^{-1}r-\frac{\mu \omega _{c}}{4\hbar }r^{2}\right]
\end{equation*}%
\begin{equation}
\times \sum_{n=0}a_{n}r^{n+\sqrt{m^{\prime }{}^{2}+\frac{2\mu r_{e}^{2}D_{e}%
}{\hbar ^{2}}}}\frac{1}{\sqrt{2\pi }}e^{im\phi },
\end{equation}%
respectively. Setting $B=0$ and $\Phi _{AB}=0,$ we obtain%
\begin{equation}
E_{nm}=-\frac{2\mu r_{e}^{2}D_{e}^{2}}{\hbar ^{2}}\left( n+\frac{1}{2}+\sqrt{%
m^{2}+\frac{2\mu r_{e}^{2}D_{e}}{\hbar ^{2}}}\right) ^{-2},
\end{equation}%
and%
\begin{equation}
\psi (\vec{r},\phi )=\exp \left[ -\frac{2\mu r_{e}D_{e}}{\hbar ^{2}}\left( n+%
\frac{1}{2}+\sqrt{m^{2}+\frac{2\mu r_{e}^{2}D_{e}}{\hbar ^{2}}}\right) ^{-1}r%
\right] \sum_{n=0}a_{n}r^{n+\sqrt{m^{2}+\frac{2\mu r_{e}^{2}D_{e}}{\hbar ^{2}%
}}}\frac{1}{\sqrt{2\pi }}e^{im\phi }.
\end{equation}%
The 3D non-relativistic energy solutions for Mie-type potential are obtained
by setting $m=l+1/2$ where $l$ is the rotational quantum number to obtain%
\begin{equation}
E_{nl}=-\frac{2\mu r_{e}^{2}D_{e}^{2}}{\hbar ^{2}}\left( n+\frac{1}{2}+\sqrt{%
\frac{2\mu r_{e}^{2}D_{e}}{\hbar ^{2}}+\left( l+\frac{1}{2}\right) ^{2}}%
\right) ^{-2},\text{ }n=0,1,2,\cdots ,\text{ }l=0,1,2,\cdots
\end{equation}%
which is identical to Eq. (30) of Ref. \cite{SM43} when $D=3$. In Table 3, for the N$_2$ amd CH molecules, when the AB flux field is set to zero and the magnetic field strength increases this result to slightly decreasing of the energy values. However its also noticed that the energy level is shifted (splitted) slightly up for $m=-1$ whereas shifted down for $m=1$. On the other hand when B is set to zero and the AB flux field changed, this flux field has a similar influence on the energy states as the magnetic field change in step 1.
\section{Concluding Remarks}

In this paper, we have carried out analysis for the 2D Schr\"{o}dinger equation with a specific potential function in the presence of external uniform magnetic $\overrightarrow{B}$ and AB flux $\Phi _{AB}$ fields by using an appropriate wave function ansatz. The restrictions on the potential parameters and $\beta $ have been given. The problem is then solved in 2D space and the bound state energy solutions are found under the influence of external magnetic and AB flux fields. The 3D bound state energy levels and wave functions are obtained in closed form to show the accuracy of the present model. As an application, we applied the energy formula on the pseudo-harmonic, harmonic, generalized Kratzer, Mie-type and Coulombic potentials models for various strength of external uniform magnetic and AB flux fields. The external fields produce splitting in the energy levels which is essentially dependent on the strength of the applied fields.

As a further application, we generated the non-relativistic energy levels of few diatomic molecules: NO, CO, N$_{2}$ and CH for various vibrational $n$ and rotational $l$ quantum numbers. The spectroscopic constant for these molecules are given in Ref. \cite{SM38} where the 3D pseudo-harmonic potential have been considered. We have developed keen interest in these molecules so as to be able to make comparism with the previous work in the absent of external magnetic field. The criterion for selecting these diatomic molecules is nothing but for their intermediate importance in the chemical physics, chemical industry, biological sciences and related areas. NO is an important messenger molecule involved in many pathological and physiological processes within the mammalian body \cite{SM61}.  CO is the simplest oxocarbon and isoelectronic with the cyanide ion and molecular nitrogen. This molecule has received a great deal of attention in clinical section as a biological regulator. N$_2$ molecules is very essential in indusry for production of ammonia, nitric acid and explosives. The emission spectrum of CH can be used to understand double spectral of diatomic molecules.

The energy levels of these diatomic molecules have been calculated in 2D space under the influence of magnetic field and Aharonov-Bohm field. It is noticed that the magnetic field has a
small influence on the energy levels when $m=0$. However, when $m\neq 0$, it removes the degeneracy and makes a narrow shift for the energy states. On the other hand, the effect of the Aharonov-Bohm field is greater as it creates a wider shift for $m\neq 0$ and its influence on $m=0$ states is found to be greater than that of the magnetic field. The $m\neq 0$ states are much more sensitive to the two fields.

\section*{Acknowledgments}
We thank the kind referee for the positive enlightening comments and suggestions, which have greatly helped us in making improvements to this paper.

\begin{landscape}
\begin{table}[tbp]
{\small
\caption{The rovibrational energy levels of the harmonic oscillator for NO, CO, N$_{2}$ and CH diatomic molecules with various vibrational $n$ and rotational $l$ quantum states for $D=3.$}
\vspace*{10pt}{
\begin{tabular}{cccccccccc}\hline\hline
{}&{}&{}&{}&{}&{}&{}&{}&{}&{}\\[-1.0ex]
$n$ & $l$& $E_{nl}^{(\text{NO})}$ $(eV)$ & Ref. \cite{SM38} & $E_{nl}^{(\text{CO})}$ $(eV)$ & Ref. \cite{SM38} & $E_{nl}^{(\text{N}_{2})}$ $(eV)$ & Ref. \cite{SM38} & $E_{nl}^{(\text{CH})}$ $(eV)$ & Ref. \cite{SM38}\\[2ex]\hline\hline
0 & 0 & 0.08251026 & 0.0824883 & 0.10195779 & 0.1019306 & 0.10918501 & 0.1091559 & 0.16867933 & 0.1686344 \\ 
1 & 0 & 0.24742511 & 0.2473592 & 0.30575368 & 0.3056722 & 0.32743034 & 0.3273430 & 0.50514181 & 0.5050072 \\ 
1 & 1 & 0.24784775 & 0.2477817 & 0.30623238 & 0.3061508 & 0.32792905 & 0.3278417 & 0.50872564 & 0.5085903 \\ 
2 & 0 & 0.41233997 & 0.4122301 & 0.50954957 & 0.5094137 & 0.54567566 & 0.5455302 & 0.84160429 & 0.8413800 \\ 
2 & 1 & 0.41276261 & 0.4126526 & 0.51002827 & 0.5098923 & 0.54617437 & 0.5460288 & 0.84518812 & 0.8449631 \\ 
2 & 2 & 0.41360783 & 0.4134977 & 0.51098564 & 0.5108495 & 0.54717177 & 0.5470260 & 0.85235089 & 0.8521246 \\ 
3 & 0 & 0.57725482 &           & 0.71334546 &           & 0.76392098 &           & 1.17806677 &           \\
 
3 & 1 & 0.57767747 &           & 0.71382416 &           & 0.76441969 &           & 1.18165060 & \\ 
3 & 2 & 0.57852269 &           & 0.71478153 &           & 0.76541710 &           & 1.18881337 & \\ 
3 & 3 & 0.57979045 &           & 0.71621750 &           & 0.76691313 &           & 1.19954539 & \\ 
4 & 0 & 0.74216968 & 0.7419718 & 0.91714135 & 0.9168969 & 0.98216631 & 0.9819045 & 1.51452925 & 1.5141255 \\ 
4 & 1 & 0.74259232 & 0.7423944 & 0.91762005 & 0.9173755 & 0.98266502 & 0.9824031 & 1.51811307 & 1.5177087 \\ 
4 & 2 & 0.74343754 & 0.7432395 & 0.91857742 & 0.9183327 & 0.98366242 & 0.9834003 & 1.52527585 & 1.5248701 \\ 
4 & 3 & 0.74470531 & 0.7445070 & 0.92001339 & 0.9197684 & 0.98515845 & 0.9848961 & 1.53600786 & 1.5356002 \\ 
4 & 4 & 0.74639552 & 0.7461969 & 0.92192787 & 0.9216825 & 0.98715302 & 0.9868903 & 1.55029463 & 1.5498843 \\ 
5 & 0 & 0.90708453 & 0.9068427 & 1.12093724 & 1.1206384 & 1.20041163 & 1.2000916 & 1.85099173 & 1.8504983 \\ 
5 & 1 & 0.90750718 & 0.9072653 & 1.12141594 & 1.1211170 & 1.20091034 & 1.2005902 & 1.85457555 & 1.8540615 \\ 
5 & 2 & 0.90835240 & 0.9081104 & 1.12237331 & 1.1220742 & 1.20190774 & 1.2015875 & 1.86173833 & 1.8612429 \\ 
5 & 3 & 0.90962017 & 0.9093779 & 1.12380928 & 1.1235099 & 1.20340377 & 1.2030832 & 1.87247034 & 1.8719729 \\ 
5 & 4 & 0.91131038 & 0.9110678 & 1.12572376 & 1.1254240 & 1.20539834 & 1.2050774 & 1.88675711 & 1.8862571 \\ 
5 & 5 & 0.91342289 & 0.9131799 & 1.12811660 & 1.1278165 & 1.20789131 & 1.2075699 & 1.90457946 & 1.9040761 \\ 
\hline\hline
\end{tabular}}
\vspace*{-1pt}}
\end{table}
\end{landscape}

\begin{table}[tbp]
{\small
\caption{The rovibrational energy levels of the pseudoharmonic oscillator for N$_{2}$ and CH diatomic molecules with various $n$ and $m$ quantum states for $D=2$ in the presence and absence of external magnetic field, B and AB flux field, $\xi$.}\vspace*{10pt}{
\begin{tabular}{cccccc}\hline\hline
{}&{}&{}&{}&{}&{}\\[-1.0ex]
$E_{nm} (eV)$ &  & N$_{2}$ &  &  &  \\[1ex]\hline
$n$	&$m$	&$E_{nm}$$(\xi=0, B=0)$	&$E_{nm}$$(\xi=0, B=1)$	&$E_{nm}$$(\xi=0, B=2)$	&$E_{nm}$$(\xi=0, B=3)$ \\[1ex]\hline 
0 & \ 0 & 0.10912268 & 0.10912268 & 0.10912271 & 0.10912275 \\ 
1 & \ 0 & 0.32736800 & 0.32736800 & 0.32736804 & 0.32736807 \\ 
  & \ 1 & 0.32761735 & 0.32762019 & 0.32762307 & 0.32762595 \\ 
  &  -1 & 0.32761735 & 0.32761451 & 0.32761169 & 0.32760889 \\ 
2 & \ 0 & 0.54561333 & 0.54561333 & 0.54561336 & 0.54561340 \\ 
  & \ 1 & 0.54586267 & 0.54586551 & 0.54586840 & 0.54587127 \\ 
  &  -1 & 0.54586267 & 0.54585983 & 0.54585702 & 0.54585421 \\ 
3 & \ 0 & 0.76385865 & 0.76385865 & 0.76385869 & 0.76385872 \\ 
  & \ 1 & 0.76410800 & 0.76411084 & 0.76411372 & 0.76411659 \\ 
  &  -1 & 0.76410800 & 0.76410516 & 0.76410234 & 0.76409953 \\ \hline
$n$	&$m$	&$E_{nm}$$(\xi=0, B=0)$	&$E_{nm}$$(\xi=1, B=0)$	&$E_{nm}$$(\xi=2, B=0)$	&$E_{nm}$$(\xi=3, B=0)$\\[1ex]\hline 
0 & \ 0 & 0.10912268 & 0.10937203 & 0.11012011 & 0.11136681 \\ 
1 & \ 0 & 0.32736800 & 0.32761735 & 0.32836543 & 0.32961213 \\ 
  & \ 1 & 0.32761735 & 0.32836543 & 0.32961213 & 0.33135744 \\ 
  &  -1 & 0.32761735 & 0.32736800 & 0.32761735 & 0.32836543 \\ 
2 & \ 0 & 0.54561333 & 0.54586267 & 0.54661075 & 0.54785746 \\ 
  & \ 1 & 0.54586267 & 0.54661075 & 0.54785746 & 0.54960277 \\ 
  &  -1 & 0.54586267 & 0.54561333 & 0.54586267 & 0.54661075 \\ 
3 & \ 0 & 0.76385865 & 0.76410800 & 0.76485608 & 0.76610278 \\ 
  & \ 1 & 0.76410800 & 0.76485608 & 0.76610278 & 0.76784809 \\ 
  &  -1 & 0.76410800 & 0.76385865 & 0.76410800 & 0.76485608 \\ 
		\hline
$E_{nm}$ $(eV)$ &  & CH &  &  & \\[1ex]\hline
$n$	&$m$	&$E_{nm}$$(\xi=0, B=0)$	&$E_{nm}$$(\xi=0, B=1)$	&$E_{nm}$$(\xi=0, B=2)$	&$E_{nm}$$(\xi=0, B=3)$ \\[1ex]\hline 
0 & \ 0 & 0.16823124 & 0.16823130 & 0.16823150 & 0.16823182 \\ 
1 & \ 0 & 0.50469372 & 0.50469378 & 0.50469399 & 0.50469433 \\ 
  & \ 1 & 0.50648593 & 0.50650742 & 0.50652905 & 0.50655081 \\ 
  & -1& 0.50648593 & 0.50646458 & 0.50644336 & 0.50642228 \\ 
2 & \ 0 & 0.84115620 & 0.84115627 & 0.84115648 & 0.84115683 \\ 
  & \ 1 & 0.84294841 & 0.84296991 & 0.84299154 & 0.84301332 \\ 
  &  -1& 0.84294841 & 0.84292706 & 0.84290585 & 0.84288478 \\ 
3 & \ 0 & 1.17761867 & 1.17761875 & 1.17761897 & 1.17761934 \\ 
	& \ 1 & 1.17941089 & 1.17943239 & 1.17945403 & 1.17947582 \\ 
	& -1& 1.17941089 & 1.17938954 & 1.17936834 & 1.17934729 \\ \hline
$n$	&$m$	&$E_{nm}$$(\xi=0, B=0)$	&$E_{nm}$$(\xi=1, B=0)$	&$E_{nm}$$(\xi=2, B=0)$	&$E_{nm}$$(\xi=3, B=0)$\\[1ex]\hline 
0 & \ 0 & 0.16823124 & 0.17002346 & 0.17539767 & 0.18434658 \\ 
1 & \ 0 & 0.50469372 & 0.50648593 & 0.51186015 & 0.52080906 \\ 
  & \ 1 & 0.50648593 & 0.51186015 & 0.52080906 & 0.53332056 \\ 
	&  -1 & 0.50648593 & 0.50469372 & 0.50648593 & 0.51186015 \\ 
2 & \ 0 & 0.84115620 & 0.84294841 & 0.84832263 & 0.85727154 \\ 
	& \ 1 & 0.84294841 & 0.84832263 & 0.85727154 & 0.86978304 \\ 
	&  -1 & 0.84294841 & 0.84115620 & 0.84294841 & 0.84832263 \\ 
3 & \ 0 & 1.17761867 & 1.17941089 & 1.18478511 & 1.19373402 \\ 
	& \ 1 & 1.17941089 & 1.18478511 & 1.19373402 & 1.20624552 \\ 
	&  -1 & 1.17941089 & 1.17761867 & 1.17941089 & 1.18478511 \\ 
\hline
\end{tabular}}
\vspace*{-1pt}}
\end{table}

\begin{table}[tbp]
{\small
\caption{The rovibrational energy levels of the Kratzer-Fues potential for N$_{2}$ and CH diatomic molecules with various vibrational $n$ and rotational $l\ $quantum states for $D=3.$}\vspace*{10pt}{
\begin{tabular}{cccc}\hline\hline
{}&{}&{}&{}\\[-1.0ex]
$-E_{nm} (eV)$ &  &  &   \\[1ex]\hline
$n$	&$m$	&N$_{2}$	&CH \\[1ex]\hline 
$0$ & $0$ & $11.88375702$ & $3.86419448$ \\ 
$1$ & $0$ & $11.77611669$ & $3.70626716$ \\ 
$1$ & $1$ & $11.77562811$ & $3.70300883$ \\ 
$2$ & $0$ & $11.66993223$ & $3.55782724$ \\ 
$2$ & $1$ & $11.66945025$ & $3.55476264$ \\ 
$2$ & $2$ & $11.66848639$ & $3.54864946$ \\ 
$3$ & $0$ & $11.56517751$ & $3.41812972$ \\ 
$3$ & $1$ & $11.56470201$ & $3.41524380$ \\ 
$3$ & $2$ & $11.56375110$ & $3.40948682$ \\ 
$3$ & $3$ & $11.56232502$ & $3.40088832$ \\ 
$4$ & $0$ & $11.46182698$ & $3.28650134$ \\ 
$4$ & $1$ & $11.46135783$ & $3.28378047$ \\ 
$4$ & $2$ & $11.46041964$ & $3.27835254$ \\ 
$4$ & $3$ & $11.45901264$ & $3.27024501$ \\ 
$4$ & $4$ & $11.45713719$ & $3.25949862$ \\ 
$5$ & $0$ & $11.35985565$ & $3.16233240$ \\ 
$5$ & $1$ & $11.35939275$ & $3.15976423$ \\ 
$5$ & $2$ & $11.35846705$ & $3.15464074$ \\ 
$5$ & $3$ & $11.35707878$ & $3.14698750$ \\ 
$5$ & $4$ & $11.35522830$ & $3.13684243$ \\ 
$5$ & $5$ & $11.35291603$ & $3.12425536$ \\ \hline
\end{tabular}}
\vspace*{-1pt}}
\end{table}
\begin{table}[tbp]
{\small
\caption{The rovibrational energy levels of the Kratzer potential for N$_{2}$ and CH diatomic molecule with various $n$ andl $m$ quantum states for $D=2$ in the presence and absence of external magnetic field, B and AB flux field, $\xi$ .}\vspace*{10pt}{
\begin{tabular}{cccccc}\hline\hline
{}&{}&{}&{}&{}&{}\\[-1.0ex]
$-E_{nm}$ $(eV)$ &  & N$_{2}$ &  &  &  \\[1ex]\hline 
$n$	&$m$	&$E_{nm}$$(\xi=0, B=0)$	&$E_{nm}$$(\xi=0, B=1)$	&$E_{nm}$$(\xi=0, B=2)$	&$E_{nm}$$(\xi=0, B=3)$ \\[1ex]\hline 
0 & \ 0 & 11.88381894 & 11.88371512 & 11.88361129 & 11.88350747 \\ 
1 & \ 0 & 11.77617777 & 11.77607347 & 11.77596918 & 11.77586488 \\ 
  & \ 1 & 11.77593347 & 11.77582870 & 11.77572393 & 11.77561916 \\ 
  &  -1 & 11.77593347 & 11.77582965 & 11.77572582 & 11.77562200 \\ 
2 & \ 0 & 11.66999248 & 11.66988771 & 11.66978294 & 11.66967818 \\ 
  & \ 1 & 11.66975149 & 11.66964625 & 11.66954101 & 11.66943577 \\ 
  &  -1 & 11.66975149 & 11.66964719 & 11.66954290 & 11.66943860 \\ 
3 & \ 0 & 11.56523696 & 11.56513172 & 11.56502648 & 11.56492124 \\ 
  & \ 1 & 11.56499920 & 11.56489349 & 11.56478777 & 11.56468206 \\ 
  &  -1 & 11.56499920 & 11.56489443 & 11.56478966 & 11.56468489 \\ \hline 
$n$	&$m$	&$E_{nm}$$(\xi=0, B=0)$	&$E_{nm}$$(\xi=1, B=0)$	&$E_{nm}$$(\xi=2, B=0)$	&$E_{nm}$$(\xi=3, B=0)$\\[1ex]\hline 
0 & \ 0 & 11.88381894 & 11.88357129 & 11.88282838 & 11.88159041 \\ 
1 & \ 0 & 11.77617777 & 11.77593347 & 11.77520064 & 11.77397945 \\ 
  & \ 1 & 11.77593347 & 11.77520064 & 11.77397945 & 11.77227022 \\ 
	&  -1 & 11.77593347 & 11.77617777 & 11.77593347 & 11.77520064 \\ 
2 & \ 0 & 11.66999248 & 11.66975149 & 11.66902854 & 11.66782383 \\ 
	& \ 1 & 11.66975149 & 11.66902854 & 11.66782383 & 11.66613766 \\ 
	&  -1 & 11.66975149 & 11.66999248 & 11.66975149 & 11.66902854 \\ 
3 & \ 0 & 11.56523696 & 11.56499920 & 11.56428596 & 11.56309744 \\ 
	& \ 1 & 11.56499920 & 11.56428596 & 11.56309744 & 11.56143392 \\ 
	&  -1 & 11.56499920 & 11.56523696 & 11.56499920 & 11.56428596 \\ \hline
$-E_{nm}$ $(eV)$ &  & CH &  &  &  \\[1ex]\hline 
$n$	&$m$	&$E_{nm}$$(\xi=0, B=0)$	&$E_{nm}$$(\xi=0, B=1)$	&$E_{nm}$$(\xi=0, B=2)$	&$E_{nm}$$(\xi=0, B=3)$ \\[1ex]\hline 
0 & \ 0 & 3.86462852 & 3.86445802 & 3.86428753 & 3.86411704 \\ 
1 & \ 0 & 3.70667486 & 3.70650081 & 3.70632676 & 3.70615271 \\ 
  & \ 1 & 3.70504461 & 3.70486696 & 3.70468932 & 3.70451167 \\ 
	&  -1 & 3.70504461 & 3.70487408 & 3.70470355 & 3.70453302 \\ 
2 & \ 0 & 3.55821069 & 3.55803309 & 3.55785548 & 3.55767787 \\ 
	& \ 1 & 3.55667739 & 3.55649619 & 3.55631498 & 3.55613378 \\ 
	&  -1 & 3.55667739 & 3.55650330 & 3.55632921 & 3.55615512 \\ 
3 & \ 0 & 3.41849082 & 3.41830965 & 3.41812848 & 3.41794732 \\ 
	& \ 1 & 3.41704692 & 3.41686216 & 3.41667740 & 3.41649264 \\ 
	&  -1 & 3.41704692 & 3.41686928 & 3.41669163 & 3.41651398 \\ \hline
$n$	&$m$	&$E_{nm}$$(\xi=0, B=0)$	&$E_{nm}$$(\xi=1, B=0)$	&$E_{nm}$$(\xi=2, B=0)$	&$E_{nm}$$(\xi=3, B=0)$\\[1ex]\hline  
0 & \ 0 & 3.86462852 & 3.86289297 & 3.85769569 & 3.84906466 \\ 
1 & \ 0 & 3.70667486 & 3.70504461 & 3.70016252 & 3.69205445 \\ 
  & \ 1 & 3.70504461 & 3.70016252 & 3.69205445 & 3.68076315 \\ 
  &  -1 & 3.70504461 & 3.70667486 & 3.70504461 & 3.70016252 \\ 
2 & \ 0 & 3.55821069 & 3.55667739 & 3.55208550 & 3.54445900 \\ 
  & \ 1 & 3.55667739 & 3.55208550 & 3.54445900 & 3.53383747 \\ 
  &  -1 & 3.55667739 & 3.55821069 & 3.55667739 & 3.55208550 \\ 
3 & \ 0 & 3.41849082 & 3.41704692 & 3.41272269 & 3.40554035 \\ 
  & \ 1 & 3.41704692 & 3.41272269 & 3.40554035 & 3.39553666 \\ 
  &  -1 & 3.41704692 & 3.41849082 & 3.41704692 & 3.41272269 \\ \hline
\end{tabular}}
\vspace*{-1pt}}
\end{table}

\begin{table}[tbp]
{\small
\caption{The rovibrational energy levels of the Mie-type potential for N$_{2}$ and CH diatomic molecule with various $n$ andl $m$ quantum states for $D=2$ in the presence and absence of external magnetic field, B and AB flux field, $\xi$ .}\vspace*{10pt}{
\begin{tabular}{cccccc}\hline\hline
{}&{}&{}&{}&{}&{}\\[-1.0ex]
$-E_{nm}$ $(eV)$ &  & N$_{2}$ &  &  &  \\[1ex]\hline 
$n$	&$m$	&$E_{nm}$$(\xi=0, B=0)$	&$E_{nm}$$(\xi=0, B=1)$	&$E_{nm}$$(\xi=0, B=2)$	&$E_{nm}$$(\xi=0, B=3)$ \\[1ex]\hline 
0	&\ 0	&11.88381894	&11.88371512	&11.88361129	&11.88350747\\
1	&\ 0	&11.77617777	&11.77607347	&11.77596918	&11.77586488\\
	&\ 1	&11.77593347	&11.77582870	&11.77572393	&11.77561916\\
	&-1	  &11.77593347	&11.77582965	&11.77572582	&11.77562200\\
2	&\ 0	&11.66999248	&11.66988771	&11.66978294	&11.66967818\\
	&\ 1	&11.66975149	&11.66964625	&11.66954101	&11.66943577\\
	&-1  	&11.66975149	&11.66964719	&11.66954290	&11.66943860\\
3	&\ 0	&11.56523696	&11.56513172	&11.56502648	&11.56492124\\
	&\ 1	&11.56499920	&11.56489349	&11.56478777	&11.56468206\\
	&-1	  &11.56499920	&11.56489443	&11.56478966	&11.56468489\\ \hline 
$n$	&$m$	&$E_{nm}$$(\xi=0, B=0)$	&$E_{nm}$$(\xi=1, B=0)$	&$E_{nm}$$(\xi=2, B=0)$	&$E_{nm}$$(\xi=3, B=0)$\\[1ex]\hline 
0	&\ 0	&11.88381894	&11.88357129	&11.88282838	&11.88159041\\
1	&\ 0	&11.77617777	&11.77593347	&11.77520064	&11.77397945\\
	&\ 1	&11.77593347	&11.77520064	&11.77397945	&11.77227022\\
	&-1	  &11.77593347	&11.77617777	&11.77593347	&11.77520064\\
2	&\ 0	&11.66999248	&11.66975149	&11.66902854	&11.66782383\\
	&\ 1	&11.66975149	&11.66902854	&11.66782383	&11.66613766\\
	&-1 	&11.66975149	&11.66999248	&11.66975149	&11.66902854\\
3	&\ 0	&11.56523696	&11.56499920	&11.56428596	&11.56309744\\
	&\ 1	&11.56499920	&11.56428596	&11.56309744	&11.56143392\\
	&-1 	&11.56499920	&11.56523696	&11.56499920	&11.56428596\\ \hline 
	$-E_{nm}$ $(eV)$ &  & CH &  &  &  \\[1ex]\hline 
$n$	&$m$	&$E_{nm}$$(\xi=0, B=0)$	&$E_{nm}$$(\xi=0, B=1)$	&$E_{nm}$$(\xi=0, B=2)$	&$E_{nm}$$(\xi=0, B=3)$\\[1ex]\hline 
0	&\ 0	&3.86462852	&3.86445802	&3.86428753	&3.86428753\\
1	&\ 0	&3.70667486	&3.70650081	&3.70632676	&3.70632676\\
	&\ 1	&3.70504461	&3.70486696	&3.70468932	&3.70468932\\
	&-1 	&3.70504461	&3.70487408	&3.70470355	&3.70470355\\
2	&\ 0	&3.55821069	&3.55803309	&3.55785548	&3.55785548\\
	&\ 1	&3.55667739	&3.55649619	&3.55631498	&3.55631498\\
	&-1	  &3.55667739	&3.55650330	&3.55632921	&3.55632921\\
3	&\ 0	&3.41849082	&3.41830965	&3.41812848	&3.41812848\\
	&\ 1	&3.41704692	&3.41686216	&3.41667740	&3.41667740\\
	&-1 	&3.41704692	&3.41686928	&3.41669163	&3.41669163\\ \hline 
$n$	&$m$	&$E_{nm}$$(\xi=0, B=0)$	&$E_{nm}$$(\xi=1, B=0)$	&$E_{nm}$$(\xi=2, B=0)$	&$E_{nm}$$(\xi=3, B=0)$\\[1ex]\hline 
0	&\ 0	&3.86462852	&3.86289297	&3.85769569	&3.84906466\\
1	&\ 0	&3.70667486	&3.70504461	&3.70016252	&3.69205445\\
	&\ 1	&3.70504461	&3.70016252	&3.69205445	&3.68076315\\
	&-1	  &3.70504461	&3.70667486	&3.70504461	&3.70016252\\
2	&\ 0	&3.55821069	&3.55667739	&3.55208550	&3.54445900\\
	&\ 1	&3.55667739	&3.55208550	&3.54445900	&3.53383747\\
	&-1	  &3.55667739	&3.55821069	&3.55667739	&3.55208550\\
3	&\ 0	&3.41849082	&3.41704692	&3.41272269	&3.40554035\\
	&\ 1	&3.41704692	&3.41272269	&3.40554035	&3.39553666\\
	&-1 	&3.41704692	&3.41849082	&3.41704692	&3.41272269\\ \hline 
\end{tabular}}
\vspace*{-1pt}}
\end{table}
\end{document}